\documentstyle[twoside,epsfig]{article}

\catcode`\@=11
\long\def\@makefntext#1{
\protect\noindent \hbox to 3.2pt {\hskip-.9pt  
$^{{\eightrm\@thefnmark}}$\hfil}#1\hfill}		

\def\@makefnmark{\hbox to 0pt{$^{\@thefnmark}$\hss}}	
	
\def\ps@myheadings{\let\@mkboth\@gobbletwo
\def\@oddhead{\hbox{}
\rightmark\hfil\eightrm\thepage}   
\def\@oddfoot{}\def\@evenhead{\eightrm\thepage\hfil
\leftmark\hbox{}}\def\@evenfoot{}
\def\sectionmark##1{}\def\subsectionmark##1{}}

\oddsidemargin=\evensidemargin
\newcounter{sectionc}\newcounter{subsectionc}\newcounter{subsubsectionc}
\renewcommand{\section}[1] {\vspace{12pt}\addtocounter{sectionc}{1} 
\setcounter{subsectionc}{0}\setcounter{subsubsectionc}{0}\noindent 
	{\tenbf\thesectionc. #1}\par\vspace{5pt}}
\renewcommand{\subsection}[1] {\vspace{12pt}\addtocounter{subsectionc}{1} 
	\setcounter{subsubsectionc}{0}\noindent 
	{\bf\thesectionc.\thesubsectionc. {\kern1pt \bfit #1}}\par\vspace{5pt}}
\renewcommand{\subsubsection}[1] {\vspace{12pt}\addtocounter{subsubsectionc}{1}
	\noindent{\tenrm\thesectionc.\thesubsectionc.\thesubsubsectionc.
	{\kern1pt \tenit #1}}\par\vspace{5pt}}
\newcommand{\nonumsection}[1] {\vspace{12pt}\noindent{\tenbf #1}
	\par\vspace{5pt}}

\newcounter{appendixc}
\newcounter{subappendixc}[appendixc]
\newcounter{subsubappendixc}[subappendixc]
\renewcommand{\thesubappendixc}{\Alph{appendixc}.\arabic{subappendixc}}
\renewcommand{\thesubsubappendixc}
	{\Alph{appendixc}.\arabic{subappendixc}.\arabic{subsubappendixc}}

\renewcommand{\appendix}[1] {\vspace{12pt}
        \refstepcounter{appendixc}
        \setcounter{figure}{0}
        \setcounter{table}{0}
        \setcounter{lemma}{0}
        \setcounter{theorem}{0}
        \setcounter{corollary}{0}
        \setcounter{definition}{0}
        \setcounter{equation}{0}
        \renewcommand{\thefigure}{\Alph{appendixc}.\arabic{figure}}
        \renewcommand{\thetable}{\Alph{appendixc}.\arabic{table}}
        \renewcommand{\theappendixc}{\Alph{appendixc}}
        \renewcommand{\thelemma}{\Alph{appendixc}.\arabic{lemma}}
        \renewcommand{\thetheorem}{\Alph{appendixc}.\arabic{theorem}}
        \renewcommand{\thedefinition}{\Alph{appendixc}.\arabic{definition}}
        \renewcommand{\thecorollary}{\Alph{appendixc}.\arabic{corollary}}
        \renewcommand{\theequation}{\Alph{appendixc}.\arabic{equation}}
        \noindent{\tenbf Appendix#1}\par\vspace{5pt}}
\newcommand{\subappendix}[1] {\vspace{12pt}
        \refstepcounter{subappendixc}
        \noindent{\bf Appendix \thesubappendixc. {\kern1pt \bfit #1}}
	\par\vspace{5pt}}
\newcommand{\subsubappendix}[1] {\vspace{12pt}
        \refstepcounter{subsubappendixc}
        \noindent{\rm Appendix \thesubsubappendixc. {\kern1pt \tenit #1}}
	\par\vspace{5pt}}

\newcommand{\textlineskip}{\baselineskip=13pt}
\newcommand{\smalllineskip}{\baselineskip=10pt}

\def\eightcirc{
\begin{picture}(0,0)
\put(4.4,1.8){\circle{6.5}}
\end{picture}}
\def\eightcopyright{\eightcirc\kern2.7pt\hbox{\eightrm c}} 

\newcommand{\copyrightheading}[1]
	{\vspace*{-2.5cm}\smalllineskip{\flushleft
	{\footnotesize International Journal of Theoretical and Applied Finance#1}\\
	{\footnotesize $\eightcopyright$\, World Scientific Publishing
	 Company}\\
	 }}

\newcommand{\pub}[1]{{\begin{center}\footnotesize\smalllineskip 
	#1\\		
	\end{center}
	}}

\def\abstracts#1#2#3{{
	\centering{\begin{minipage}{4.5in}\baselineskip=10pt\footnotesize
	\parindent=0pt #1\par 
	\parindent=15pt #2\par
	\parindent=15pt #3
	\end{minipage}}\par}} 

\renewenvironment{thebibliography}[1]
	{\frenchspacing
	 \ninerm\baselineskip=11pt
	 \begin{list}{\arabic{enumi}.}
        {\usecounter{enumi}\setlength{\parsep}{0pt}     
	 \setlength{\leftmargin 12.7pt}{\rightmargin 0pt} 
         \setlength{\itemsep}{0pt} \settowidth
	{\labelwidth}{#1.}\sloppy}}{\end{list}}

\newcounter{itemlistc}
\newcounter{romanlistc}
\newcounter{alphlistc}
\newcounter{arabiclistc}

\newcommand{\fcaption}[1]{
        \refstepcounter{figure}
        \setbox\@tempboxa = \hbox{\footnotesize Fig.~\thefigure. #1}
        \ifdim \wd\@tempboxa > 5in
           {\begin{center}
        \parbox{5in}{\footnotesize\smalllineskip Fig.~\thefigure. #1}
            \end{center}}
        \else
             {\begin{center}
             {\footnotesize Fig.~\thefigure. #1}
              \end{center}}
        \fi}

\newcommand{\tcaption}[1]{
        \refstepcounter{table}
        \setbox\@tempboxa = \hbox{\footnotesize Table~\thetable. #1}
        \ifdim \wd\@tempboxa > 5in
           {\begin{center}
        \parbox{5in}{\footnotesize\smalllineskip Table~\thetable. #1}
            \end{center}}
        \else
             {\begin{center}
             {\footnotesize Table~\thetable. #1}
              \end{center}}
        \fi}

\def\@citex[#1]#2{\if@filesw\immediate\write\@auxout
	{\string\citation{#2}}\fi
\def\@citea{}\@cite{\@for\@citeb:=#2\do
	{\@citea\def\@citea{,}\@ifundefined
	{b@\@citeb}{{\bf ?}\@warning
	{Citation `\@citeb' on page \thepage \space undefined}}
	{\csname b@\@citeb\endcsname}}}{#1}}

\newif\if@cghi
\def\cite{\@cghitrue\@ifnextchar [{\@tempswatrue
	\@citex}{\@tempswafalse\@citex[]}}
\def\citelow{\@cghifalse\@ifnextchar [{\@tempswatrue
	\@citex}{\@tempswafalse\@citex[]}}
\def\@cite#1#2{{$\null^{#1}$\if@tempswa\typeout
	{IJCGA warning: optional citation argument 
	ignored: `#2'} \fi}}

\def\pmb#1{\setbox0=\hbox{#1}
	\kern-.025em\copy0\kern-\wd0
	\kern.05em\copy0\kern-\wd0
	\kern-.025em\raise.0433em\box0}

\def\fnt#1#2{\footnotetext{\kern-.3em
	{$^{\mbox{\scriptsize #1}}$}{#2}}}

\def\fpage#1{\begingroup
\voffset=.3in
\thispagestyle{empty}\begin{table}[b]\centerline{\footnotesize #1}
	\end{table}\endgroup}

\def\runninghead#1#2{\pagestyle{myheadings}
\markboth{{\protect\footnotesize\it{\quad #1}}\hfill}
{\hfill{\protect\footnotesize\it{#2\quad}}}}
\font\tenrm=cmr10
\font\tenit=cmti10 
\font\tenbf=cmbx10
\font\bfit=cmbxti10 at 10pt
\font\ninerm=cmr9

\font\eightrm=cmr8

\def\qed{\hbox{${\vcenter{\vbox{			
   \hrule height 0.4pt\hbox{\vrule width 0.4pt height 6pt
   \kern5pt\vrule width 0.4pt}\hrule height 0.4pt}}}$}}

\def\theequation{\thesectionc.\arabic{equation}}	

\def\wb{\widetilde{W_t}}
\def\avd#1{\overline{#1}}
\def\s{\sigma_t}
\def\a{\alpha}
\def\N{N_n}
\def\r{\rho}
\def\l{\lambda}
\def\ldit{\chi(t)}
\def\g{\gamma}

\def\avt#1{\langle#1\rangle}

\begin{document}

\runninghead{Optimal strategies for prudent investors}
{Optimal strategies for prudent investors}

\normalsize\textlineskip
\thispagestyle{empty}
\setcounter{page}{1}

\copyrightheading{}			

\vspace*{0.88truein}

\fpage{1}
\centerline{\bf OPTIMAL STRATEGIES FOR PRUDENT INVESTORS}
\vspace*{0.37truein}
\centerline{\footnotesize ROBERTO BAVIERA}
\vspace*{0.015truein}
\centerline{\footnotesize\it Dipartimento di Fisica, Universit\`a dell'Aquila,
and Istituto Nazionale Fisica della Materia}
\baselineskip=10pt
\centerline{\footnotesize\it Via Vetoio, I-67010 Coppito, L'Aquila, Italy}
\vspace*{10pt}
\centerline{\footnotesize MICHELE PASQUINI and MAURIZIO SERVA}
\vspace*{0.015truein}
\centerline{\footnotesize\it Dipartimento di Matematica, Universit\`a dell'Aquila,
and Istituto Nazionale Fisica della Materia}
\baselineskip=10pt
\centerline{\footnotesize\it Via Vetoio, I-67010 Coppito, L'Aquila, Italy}
\vspace*{10pt}
\centerline{\footnotesize ANGELO VULPIANI}
\vspace*{0.015truein}
\centerline{\footnotesize\it Dipartimento di Fisica, Universit\`a di Roma ``La Sapienza''
and Istituto Nazionale Fisica della Materia}
\baselineskip=10pt
\centerline{\footnotesize\it  P.le A. Moro 2, I-00185 Roma, Italy}
\vspace*{0.225truein}
\pub{ \today }

\vspace*{0.21truein}
\abstracts{
We consider a stochastic model of investment on an asset of 
a stock market for a prudent investor. 
She decides to buy permanent goods with a fraction $\a$ 
of the maximum amount of money owned in her life 
in order that her economic level never decreases.
The optimal strategy is obtained by maximizing the exponential growth 
rate for a fixed $\a$.
We derive analytical expressions for the typical 
exponential growth rate of the capital and its fluctuations
by solving an one-dimensional random walk with drift.
}{}{}



\vspace*{1pt}\textlineskip	
\section{Introduction}	
\vspace*{-0.5pt}
\noindent

A large number of studies on finance have the main purpose to find 
the optimal strategy for a given kind of investment 
\cite{duffie,ingersoll,markowitz,merton}.
These problems can be tackled by looking for simplified
(but non-trivial) models which are able to describe the 
observed phenomenology and which can be, eventually, solved analytically.
For an introduction to financial problems discussed
from the point of view of the theoretical physics see 
\cite{boupot,absv,maszha,galzha,breimanI,breimanII}.

The optimal strategy is usually defined as the one which
maximizes a given utility function which takes into account the risk.
The use of utility functions introduce a high degree of
arbitrariness since the particular choice depends
on the subjective aversion to risk of the investor.
This psychological arbitrariness can be removed if one considers that
the rate of growth of the capital 
is an almost sure quantity on the long run. 
Therefore,
the best strategy can only be the one which maximizes this rate
i.e. the one which maximizes the expected logarithm of the capital. 
Any other strategy ends almost surely with
an exponentially in time smaller capital
\cite{maszha,breimanI,breimanII,kelly}.

The deep understanding of the reasons for the use of logarithmic 
optimization strategy comes from the pioneering Kelly's work \cite{kelly}.
In his paper he proposes and solves 
a model where an investor uses a fraction $l$ of her capital
to buy shares of a given asset at discrete time steps.
It is assumed that the price of the shares can, at each time, 
double or vanish, so that the investor doubles or loses 
the fraction she has invested in it. 
It is assumed that the probability $p$
of doubling is larger than $1/2$, this is a reasonable assumption
since the contrary situation implies that
is better to keep the money in a free risk investment 
(the interest rate is supposed to be vanishing). 

If the investment is absolutely sure $(p = 1)$, 
of course she will invest all the capital $( l =1 )$ at each step.
In this way after $t$ steps her capital will be $2^t$ times the original one.
On the other hand, 
if the evolution of the share price is uncertain and she wants to 
maximize the expected value of her capital, she chooses the same strategy
by investing a fraction $l = 1$.
Obviously this is not the best one: in fact it is sufficient 
only one defeat to
lose everything.

Using arguments from the theory of probability, 
Kelly has shown that the correct quantity to maximize is the expected value 
of the growth rate $\ldit$ of the capital:
this quantity corresponds to the rate of transmission 
over a channel in information theory or
to the Lyapunov exponent in dynamical systems and statistical mechanics of
disordered systems. 
The value of $\ldit$ for a particular sequence of investments 
fluctuates around the expected value $\avt{\ldit}$ and 
the fluctuations go to
zero in the limit $t \to \infty$. 

If one introduces a random interest factor $r_t$ and/or a random gain
factor $v_t$ (in the original Kelly's work $r_t = 1$ and 
$v_t = 2$) the model is still trivial \
from a mathematical point of view and it is easy
to find out the optimal strategy.
This is due to the fact that the model can be written in terms
of a multiplicative random process.
For a discussion on optimal investment strategy of a multi-asset portfolio
following Kelly's approach see \cite{maszha}.

Recently Galluccio and Zhang \cite{galzha} have considered a generalization 
of the previous model, where at each time step
several kinds of investment are possible: a sure one (i.e. the bank) and 
the other ones risky (i.e. the stock market).
They assume, as Kelly does, a simple behaviour 
for the market and find the values of
the parameters which optimize the Lyapunov exponent.
This model can be written as product of independent random matrices 
and it can be treated with standard perturbative methods \cite{cpv} or by
{\it constrained annealing} \cite{serpal,pps}.

In this paper we consider the case of an investment 
where there is a diversification between 
the stock market and permanent goods. 
A permanent good, such as a house, is characterized 
(at least as a first approximation) by a 
value that does not change in time, and it is not easy to convert
in cash.
The prudent investor wants, at least, to assure herself the same economic level 
for all her life (i.e. the capital invested in permanent goods cannot decrease).
In order to reach the goal 
she decides the permanent goods must equal a fraction $\a$  of the
maximum total capital owned in the past.
Therefore the capital cannot 
be less than this threshold and 
only the remaining part can be invested in the market. 
As a consequence the model describes the capital as a stochastic variable with 
memory.

Let us briefly sum up the contents of the paper.
In section 2, we introduce our model as a modification 
of the Kelly's one 
and we give an interpretation of the parameters introduced.
Section 3 is devoted to the analytical solution of the model.
In section 4, we compare our results with the Kelly's ones.
In section 5,
we discuss some aspects of our model and its main features,
in particular we consider the possibility of 
looking for time changing strategies of investment.

\section{The model}
\noindent
A given asset in a stock market can be always modelled 
in absence of memory by
\begin{equation}
W_{t+1} = F_t ( W_t ) \,\, .\label{genprocess}
\end{equation}
where $W_t$ is the capital at discrete time $t$ and $F_t$
is a random function, i.e. at each time $t$ one chooses among
different functions according to given probabilities \cite{pst,hph}.
The simplest case is $F_t = u_t W_t$, where the $u_t$ are independent 
stochastic variables (e.g. they 
can assume only two values as in the case of Kelly).

In the Kelly's model the investor keeps untouched a fraction $1 - l$
(with $0 \le l \le 1$) of its capital, while the rest is used
to buy shares 
with two different results: 
or she doubles her investment, either she loses it.
Therefore the capital at time $t+1$ is given by
\begin{equation}
W_{t+1} =  (1 - l )  W_t + l (1 + \s ) W_t =  (1 + l \s )  W_t
\,\, . \label{kellymodel}
\end{equation}
It follows that $u_t$ can be written as 
$$
u_t =  1 + l \s 
$$
where the dichotomic random variable $\s$ 
$$
{\rm \s=} \ \left\{ 
\begin{array}{ll}
+1  &  \mbox{with probability \ \ $p$} \\
-1  &  \mbox{with probability $1-p$} \end{array} 
\right.
$$
describes the change of the share price.

The growth rate of the capital at time $t$ is
$  \ldit = {1 \over t} \ln W_t $.
This quantity is random. Nevertheless for large t, 
because of the law of large numbers, it converges almost
surely to the Lyapunov exponent 
\begin{equation}
\l \equiv \lim_{t \to \infty} \ldit  
\,\, . \label{lyapgen}
\end{equation}

Let us notice that the optimal strategy proposed by Kelly consists 
in the maximization of $\l$ (i.e. $\avt{\ln W_t }$)
and not of $\avt{ W_t }$.
Following the naive idea to maximize $\avt{ W_t }$
one has $ l = 1 $ and  $\avt{ W_t } = (2p)^t $ which is 
much larger (at large $t$) than the corresponding value 
obtained with the maximization of 
$\l$.
Nevertheless, the naive strategy is clearly wrong since 
at large $ t $ one has a probability close to $ 1 $ to lose
everything.
On the contrary the maximization of 
$ \l $ 
has a well established theoretical motivation
in the law of large numbers.
Basically one has that for almost all the realizations the
quantity $\ldit$ at large $ t $ is close to its mean value $\l$,
i.e. the quantity $\ldit$ is self-averaging. 
Then, since we are dealing with a multiplicative process, one has that 
the probability distribution of $W_t$ 
is close to the log-normal one \cite{cpv,palvul}:

\begin{equation}
P(W_t) \simeq {
{1\over{ \sqrt{2 \pi \Delta^2 t} \  W_t } } 
{\ \exp{ - (\ln W_t - \l t)^2 \over{ 2 \Delta^2 \ t}}} }
\,\, . \label{lognorm}
\end{equation}
where
\begin{equation}
\Delta^2 = \lim_{t \to \infty} t 
\avt{({1 \over t} \ln W_t - \l)^2}
\,\, . \label{vargen}
\end{equation}
The (\ref{lognorm}) holds
for small values of $\left( \ln W_t - \l t \right) / \Delta \sqrt{t}$,
on the contrary the tails depend on the details of the multiplicative process
\cite{cpv,palvul}.
Let us remark that the investor can have a 
small but finite probability to have a very low capital at time $t$.

The optimization problem consists in choosing 
the fraction $ l $ of the capital that maximizes $\l$ given 
the probability $p > 0.5$; the result is 
$l_{max} = 2 p -1$.

Our aim is to modify the Kelly's model in order that
the capital cannot become too small in any realization 
of the random sequence $\{\s \}$.
This time the investor decides to buy shares with only a part
of the capital and she uses the other part to buy permanent goods.
The value of the goods equals a fraction $\alpha$ 
of the maximum total capital she has owned in the past.

The model can be written in the form: 
\begin{equation}
W_{t+1} = \a \widetilde{W_t} + (1 + l \s ) ( W_t - \a  \widetilde{W_t})
\,\, . \label{model}
\end{equation}

where 
\begin{equation}
\widetilde{W_t} =  \max_{\{i \le t\}} {W_i} 
\,\, . \label{threshold}
\end{equation}

Let us stress
that $\a \widetilde{W_t}$ is the part of the capital kept
untouched ($W_{t+1}$ is always larger than $\a \widetilde{W_t}$), 
and $l$ is the fraction of the available part 
$W_t - \a \wb$, risked at time $t$.
In the following $\a$ will be considered a fixed parameter 
depending on the greediness (or the prudence) of the investor.
The Kelly model is recovered for $\a = 0$.

Let us remark that the optimal strategy of the model (\ref{model}) - 
(\ref{threshold})
is, from a conceptual point of view, equivalent to the optimal strategy
of the original Kelly's problem with a suitable utility function
which takes into account the prudence of the investor. 
The model is then a non-markovian process for the  single variable $W_t$. 
A typical realization of $W_t$ and $\wb$ 
is shown in Fig. 1.

\begin{figure}
\vspace*{13pt}
\begin{center}
\mbox{\psfig{file=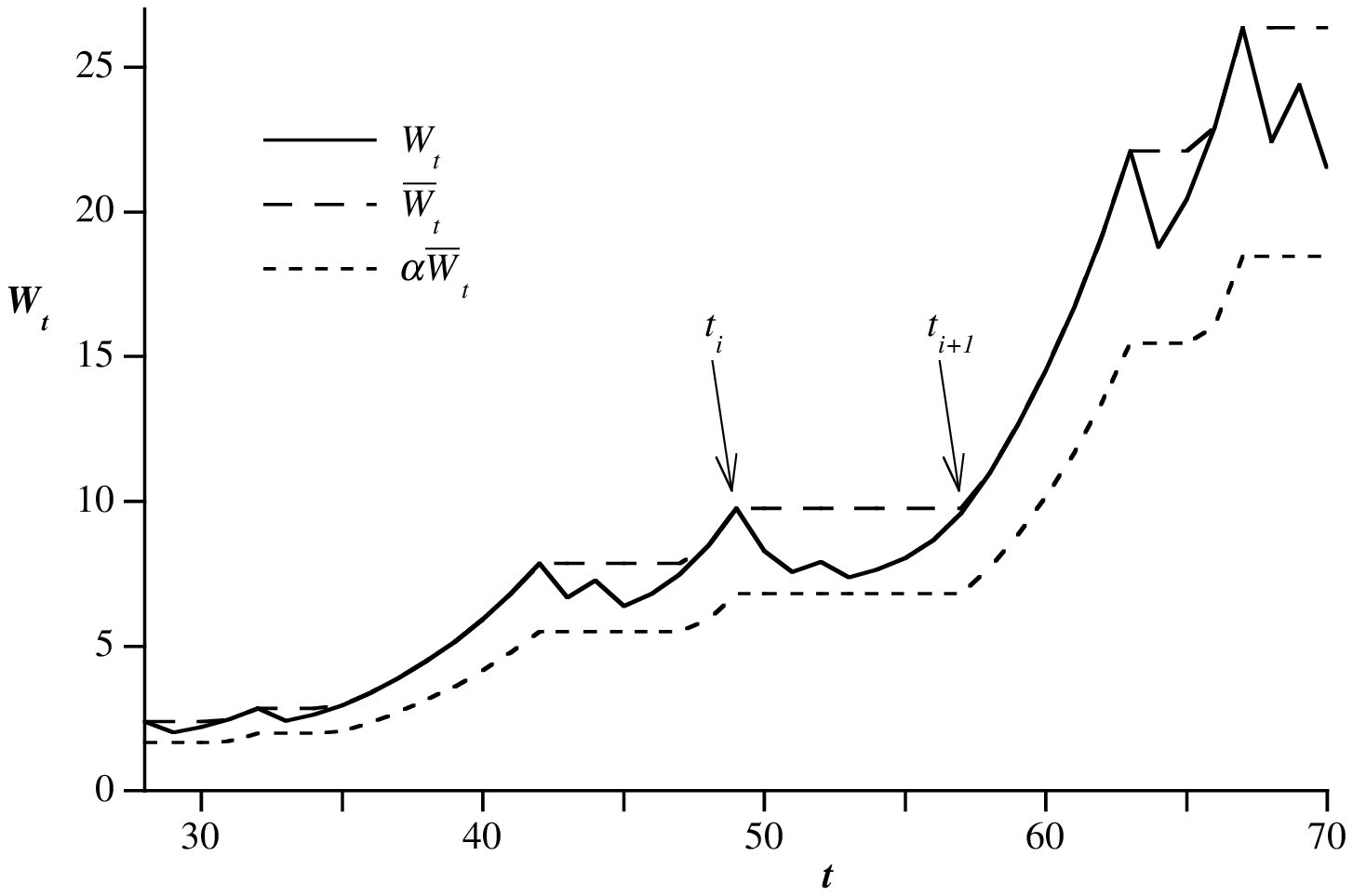,width=4.5in}}
\end{center}
\vspace*{13pt}
\fcaption{
A typical realization of the capital $W_t$ (full line), of its
maximum $\wb$ (dashed line) and of the
capital invested in permanent goods $\a \wb$ (dotted line), 
for $\a=0.7$ and $p=0.75$. The two arrows
shows a process of type (\ref{modelbetween}), where $\wb$ is constant.
}
\end{figure}

Moreover it is worth interesting that our model can be thought as 
a Markovian process of two variables $(W_t$ and $\widetilde{W_t})$ 
if we express  
the maximum capital owned in the investor's life (\ref{threshold}) as:
\begin{equation}
\widetilde{W}_{t+1} =  \max \{ \widetilde{W}_{t}, W_{t+1} \}
= \max \{ \widetilde{W}_{t},
(1 + l \s ) ( W_{t} - \a  \widetilde{W}_{t}) + \a \widetilde{W}_{t}\} 
\,\, . \label{thresholdMarkov}
\end{equation}
Equations (\ref{model}) and (\ref{thresholdMarkov}) are in fact a random map
of two variables of the form (\ref{genprocess}): 
\begin{equation}
\left( \begin{array}{ll}
W_{t+1} \\ \widetilde{W}_{t+1} \end{array} \right) = 
F_t \left( \begin{array}{ll}
W_{t} \\ \wb \end{array} \right) 
\,\, . \label{genprocesstwo}
\end{equation}
Equation (\ref{genprocesstwo}) can be thought as a product of random matrices 
of infinite order. 
This is clear if one consider the process (\ref{model}) with where now 
$\widetilde{W_t}$ is
\begin{equation}
\wb = V_t^{(1)}   = \max_{\{i = t-1, t\}} {W_i} 
\,\, . \label{threstwo}
\end{equation}
or
\begin{equation}
\wb = V_t^{(j)}   = \max_{\{t-j \le i \le t\}} {W_i} 
\,\, . \label{thresn}
\end{equation}
It is easy to realize the model (\ref{model}) with $\widetilde{W_t}$ given by
(\ref{threstwo}) can be represented 
in term of a Markov process of order 1 (i.e. the
state at time $t$ depends from the states at $t$ and $t-1$).
In a similar way using (\ref{thresn}) one has a Markov process
of order $j$.

As far as we know for a general problem like (\ref{genprocesstwo})
there is not an Oseledec theorem \cite{cpv} 
for the self-averaging of the quantity $\ldit$.
Nevertheless for our specific case we shall show in the next section that 
$\ldit$ is self-averaging.
 
\section{The solution}
\noindent
The process (\ref{model}) and  (\ref{threshold}) can be thought as 
a sequence of independent processes.
Each of them starts at time $t_i$ and ends at time $t_{i+1}$
when $\widetilde{W_t}$ changes its value
(i.e. $ W_t $ reaches a new maximum)
so that, during this time,
$\widetilde{W_t}$ is
constant and equals the starting capital $W_{t_i}$
(see Fig. 1).

Between $t_i$ and $t_{i+1}$ 
the variable $ W_t - \a W_{t_i}$ is multiplicative 
as in the Kelly's model, in fact the equation (\ref{model}) reduces to
\begin{equation}
W_{t+1} - \a W_{t_i} = (1 + l \s ) ( W_t - \a W_{t_i} ) 
\,\, . \label{modelbetween}
\end{equation}
where $W_{t_i}$ plays the role of a constant.

We notice that the final value of the $i^{th}$ process $W_{t_{i+1}}$
turns out to be proportional to its initial value $W_{t_i}$:
\begin{equation}
W_{t_{i+1}} = e^{ \g^{(i)} } W_{t_i}
\,\, . \label{modelmult}
\end{equation}
where $\g^{(i)}$ depends on all the $\{\s\}$ extracted
during the time interval $(t_i,t_{i+1})$.

This process ends when $W_{t_{i+1}}$ becomes larger than $W_{t_i}$
( $\g^{(i)} > 0 $ ) for the first time. 
Of course the time interval $N^{(i)}= t_{i+1} - t_i$ is a random quantity, 
and it depends on the  $\{\s\}$ sequence.

In this contest the global process $W_t$ (\ref{model})
 can be expressed in terms of 
$M$ independent Markovian processes $\g^{(1)}$,...,$\g^{(M)}$ as:
\begin{equation}
W_t = W_0 \prod_{i=1}^M  e^{ \g^{(i)} }
\,\, . \label{solvemodel}
\end{equation}
where $t$
equals the sum of the time duration of 
all the $M$ Markovian processes \
$t = \sum_{i=1}^M N^{(i)}$.
Let us remark that (\ref{solvemodel})
 is a product of independent random factors,
this implies that $\ldit$ reaches the value $\l$ at large $t$ for almost
all realizations.

Let us stress that the $i^{th}$ process 
described by (\ref{modelbetween}) is a one-dimensional 
random walk with positive drift $(p > 1/2)$, in terms of the
variable $\sum_{t'=t_i}^t \sigma_{t'}$.
The process ends as soon as the random walk reaches
an {\it escape point} that runs away with velocity $\r -1 \over {\r +1}$,
where $\r$ is a monotonic function of $l$ defined by
\begin{equation}
\r = - {\ln (1 -l) \over \ln (1+l)}
\,\, . \label{rhodil}
\end{equation}
with $\r \ge 1$.

Supposing that it happens with $n^{(i)}$ defeats (or negative steps
of the random walk), we find that
\begin{equation}
\g^{(i)} = \ln \left[ \a + 
(1- \a) (1-l)^{n^{(i)}} (1+l)^{N^{(i)} - n^{(i)} } \right] 
\,\, . \label{gammaeqn}
\end{equation}
Let us notice that 
$N^{(i)}$ and $n^{(i)}$ are not independent but they must satisfy
\begin{equation}
N^{(i)} = 1 + n^{(i)} + \left[ n^{(i)} \r \right] \equiv N_{n^{(i)}}
\,\, . \label{timedur}
\end{equation}
where the square bracket indicates the integer part.

Let $p(n)$ be the probability that the process $\g$ ends with $n$ defeats; it
can be written down as
\begin{equation}
p(n) = C_n (1-p)^n p^{\N -n}
\,\, . \label{distribution}
\end{equation}
where $C_n$ is the number of different ways to exit 
from the process with $n$ defeats.
The following recursive formula for $C_n$ holds (see Appendix) for $n \ge 3$:
\begin{equation}
C_n = \left(\matrix{N_{n-1} -1 \cr n-1}\right) - 
\sum_{r=1}^{n-2} \left(\matrix{N_{n-1}-N_r \cr n-r}\right) C_r
\,\, . \label{coeff}
\end{equation}
with initial values
\begin{equation}
\left\{ \begin{array}{ll} C_0 = C_1 = 1 \\
C_2 = N_1 -1 \end{array} \right.
\,\, . \label{initial}
\end{equation}
Equations (\ref{coeff}) and (\ref{initial})  represent 
a practical tool in order to numerically compute $p(n)$.

Let us notice that, fixed $p$, exists an interval of $\r$ (i.e. of $l$),
such that the $i$-th process has a non vanishing probability to have 
an infinite time duration (it happens when the mean velocity $2p-1$ 
of the random walk is lower than the velocity $\r-1 \over {\r+1}$ 
of the escape point). This implies that the growth rate $\l$ of the
capital is almost surely zero,
since its evolution remains confined in a $i$-th process, with finite $i$.
Of course it is a non optimal situation, and the following
considerations are restricted to the more interesting cases with $\l > 0$.

In the limit $M \to \infty$, 
because of the law of large numbers,
the Lyapunov exponent $\lambda$
can be written as:
\begin{equation}
\l =  \lim_{M \to \infty} { \sum^M_{i=1} \g^{(i)} \over{\sum^M_{i=1} N^{(i)} }} =
{\ \avd{\g_n} \ \over \ \avd{N_n} \ } 
\,\, . \label{lyapeqn}
\end{equation}
where the bar indicates the average according to
the distribution $p(n)$.

Let us recall that the distribution of $W_t$ is approximated by a
log-normal (\ref{lognorm}), and furthermore 
the variance (\ref{vargen}) can be written as
\begin{equation}
\Delta^2 =
 \lim_{M \to \infty}  
{\avd{\left[ 
(\sum^M_{i=1}  \g^{(i)} - \l \sum^M_{i=1} N^{(i)})^2 
\over{\sum^M_{i=1} N^{(i)}}
\right] }} 
={\avd{\left({\g_n} - \l \N \right)^2 } \over \avd{N_n}} 
\,\, . \label{vardue}
\end{equation}
where the last result is obtained simply
noticing that $ \sum^M_{i=1} N^{(i)} $ is equal, 
for the law of large numbers, to 
$ M \ \avd{N} + O(M^{1 \over{2}})$.

\section{Discussion of the results}
\noindent
As in the Kelly's model  
we maximize the Lyapunov exponent with respect to $l$,
to find the long time optimal strategy.
We have computed $\l$ and its variance using equations
(\ref{lyapeqn}) and (\ref{vardue}). The probability $p(n)$ is found out
for any $n$ smaller than
an appropriated $\widetilde{n}$ so that 
$\sum_{n=0}^{\widetilde{n}} p(n) \ge 1 - 10^{-8}$, 
$\widetilde{n}$ is typically $O(10^2)$.

\begin{figure}
\vspace*{13pt}
\begin{center}
\mbox{\psfig{file=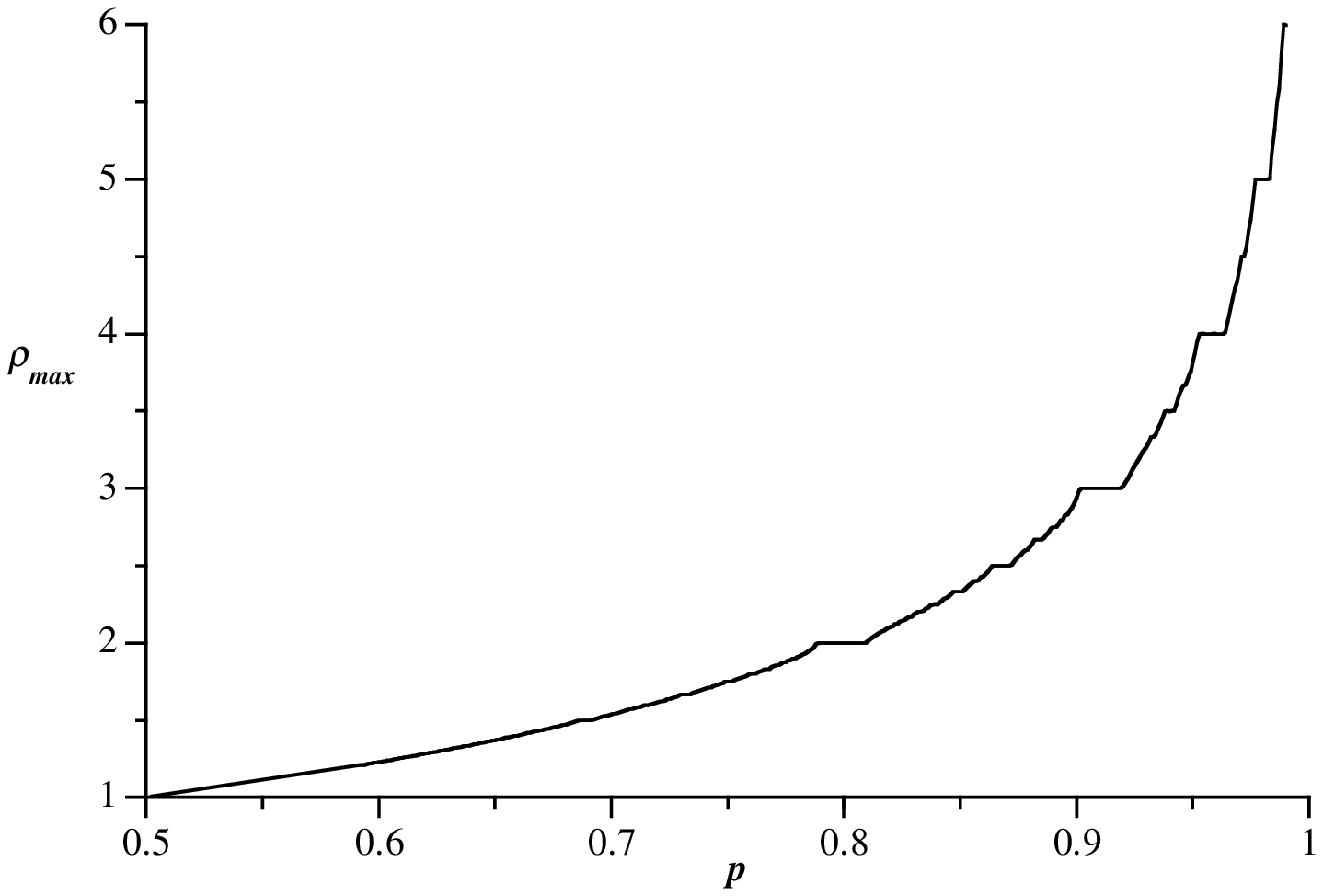,width=4.5in}}
\end{center}
\vspace*{13pt}
\fcaption{
$\r_{max}=\r(l_{max})$ (\ref{rhodil}) as function of $p$ for $\a = 0.6$.
The plateaux correspond to rational values of $\r_{max}$.
}
\end{figure}

In Fig. 2 we report $\r_{max} \equiv \r (l_{max})$ as a function of $p$ for 
$\a=0.6$. 
We observe that $\r_{max}$
is constant for some intervals of $p$. 
These plateaux correspond to rational values of  $\r_{max}$.
 This feature of the model can be explained noting that
the probability $p(n)$ 
is discontinuous around any rational $\r$.
As a consequence, the Lyapunov exponent $\l$ as function of $\r$ 
for a fixed value of $p$, has a cusp
at any rational value of $\r$. 
One of these cusps is a maximum of $\l$ in correspondence of the plateau of $p$
(see, for instance, Fig. 3a).

In this context, varying $p$ one has only a rotation of the cusp,
so that at different $p$ corresponds the same value of $\r_{max}$ that 
maximizes the Lyapunov exponent $\l$,
while for $p$ out of the plateau one has a decreasing or an 
increasing cusp, like in Fig. 3b.
The width of the plateau depends on $\a$ and it becomes larger
when $\a$ increases and when $\r$ is integer.

\begin{figure}
\vspace*{13pt}
\mbox{\centerline \tenbf a}
\begin{center}
\mbox{\psfig{file=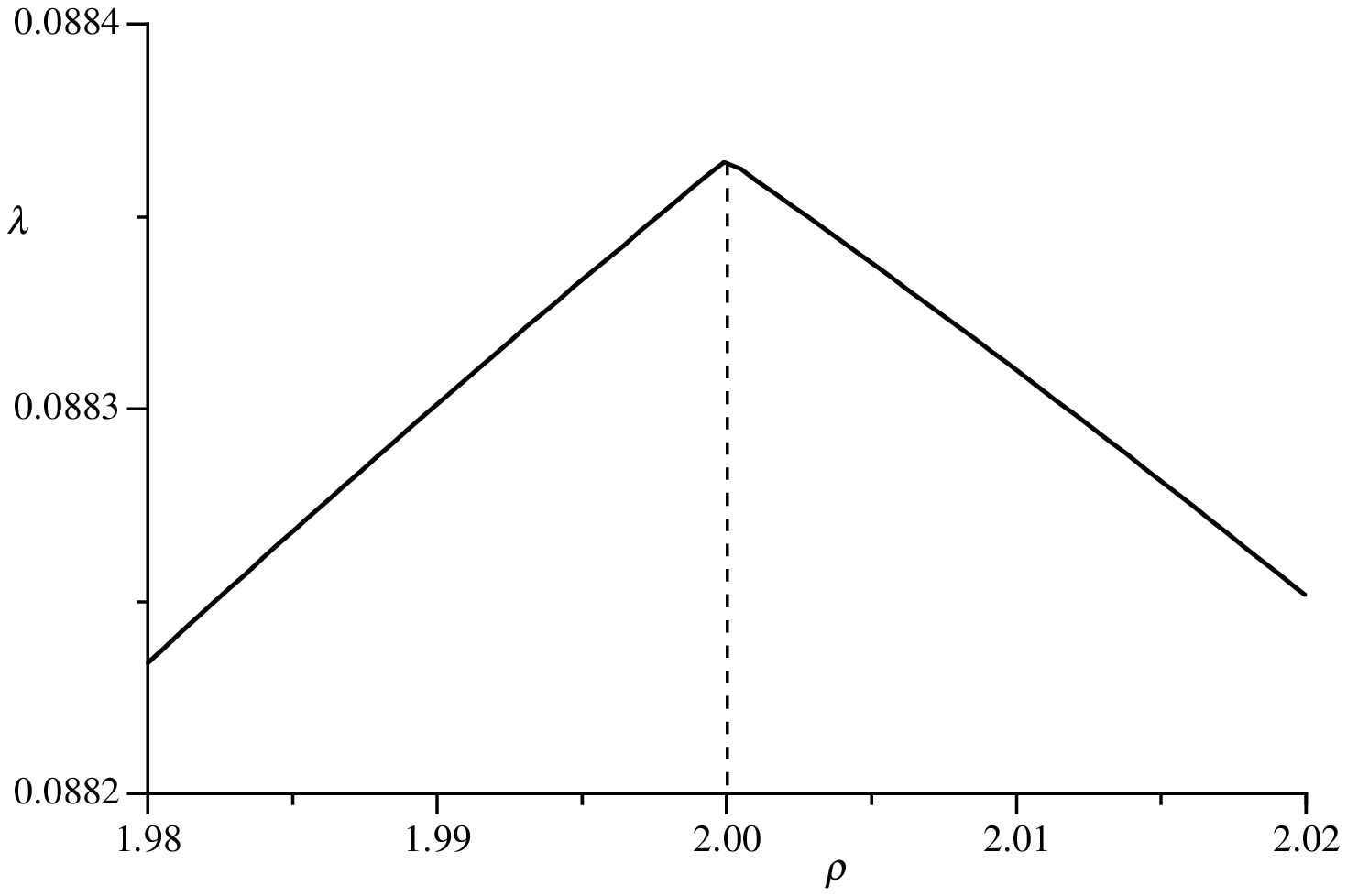,width=4in}}
\end{center}
\vspace*{26pt}
\mbox{\centerline \tenbf b}
\begin{center}
\mbox{\psfig{file=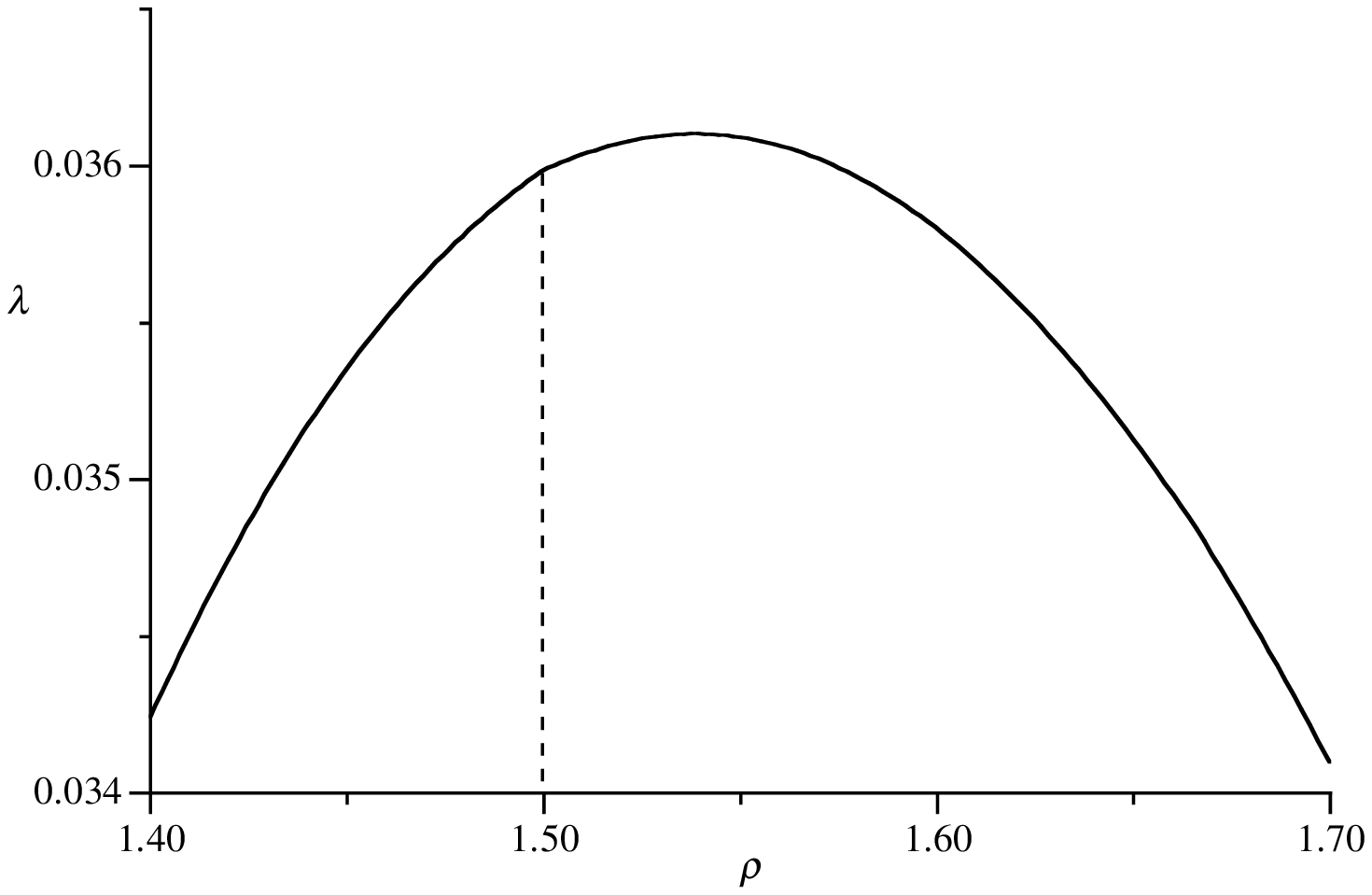,width=4in}}
\end{center}
\vspace*{13pt}
\fcaption{
Lyapunov exponent $\l$ as function of $\r$ for: \\
a) $p=0.8$ and $\a=0.6$, where the maximum is a cusp ($\r=2$); \\
b) $p=0.6$ and $\a=0.6$, where an increasing cusp ($\r=1.5$) 
is present but not in correspondence of the maximum.
}
\end{figure}

Let us restrict to the case of integer $\r$.
In order 
to simplify the notation, we use $\hat{\cdot}$ to indicate
a quantity computed at integer values of $\r$.
If we study the probability distribution of the defeats (\ref{distribution})
for fixed values of $\a $ and $p$ as a function of $\r$
we notice that,
when $\r$ crosses an integer value $\hat{\r}$,
the time durations $\{N_n\}$ (\ref{timedur}) and 
the coefficients $\{C_n\}$ (\ref{coeff})  change for every $n$.
$N_n$ changes since it depends from the integer part of $ n \r $, and
$C_n$ since the number of paths of the random walk
ending with $n$ losses depends on the cases with less defeats. 
For the same reason both of them  remain constant 
immediately on the right and on the left of $\hat{\r}$, 
at least till $n$ is big enough to have negligible effects on the
probability distribution $p(n)$.

Following this idea we perform a linear expansion in
$l$ of $\l$ around its value on the cusp.
Then we write
$$
\l^{\pm}=\l^{\pm}_0 + \l^{\pm}_1 \delta l
$$
where $\l^{+}$ and $\l^{-}$ are respectively the right and
the left limit for $l \to \hat{l} \pm 0$ of  $\l$.
After some algebra one obtains
\begin{equation}
\l^{+}_0=\l^{-}_0=\hat{\l}=\left( p- \hat{\r} (1-p) \right)
\ln(\a + (1 - \a) (1+\hat{l}))
\,\, . \label{lambdazero}
\end{equation}
and 
\begin{equation}
\l^{-}_1 = {1-\a \over 1 - \hat{l}^2} 
 \left( p 
[ 2 + \hat{b} (1 + \hat{\r}) ] - 
[ 1 + \hat{l} + \hat{\r}  \hat{b} ] \right)   
\,\, . \label{lambdaoneminus}
\end{equation}
\begin{equation}
\l^{+}_1 =
{1-\a \over \a \hat{l} (1 - \hat{l})^2} \hat{b}
[2 p - (1 + \hat{l})]
\,\, . \label{lambdaoneplus}
\end{equation}
where
\begin{equation}
\hat{b}= { \a \hat{l} (1 - \hat{l}) \over{
 \a + (1 - \a ) (1 + \hat{l}) } }
\,\, . \label{bhat}
\end{equation}
The signs of $\l^{\pm}_1$ tell us when the cusp is
a maximum of $\l$ as a function of $l$ 
(i.e. when $\l^{-}_1 \ge 0$ and $\l^{+}_1 \le 0$) .
From (\ref{lambdaoneminus}) and (\ref{lambdaoneplus}) is easy 
to see that this happens 
when $p$ is between $p_{min}$ and $p_{max}$ where
\begin{equation}
p_{min}= {  1 + \hat{l} + \hat{\r}  \hat{b} 
\over{ 2 + \hat{b} (1 + \hat{\r}) } }
\,\, . \label{pmin}
\end{equation}
\begin{equation}
p_{max}= { 1 + \hat{l} \over{2} }
\,\, . \label{pmax}
\end{equation}
Some of the widths of these plateaux (i.e.  $p_{max} - p_{min}$)
are plotted in Fig. 4.

\begin{figure}
\vspace*{13pt}
\begin{center}
\mbox{\psfig{file=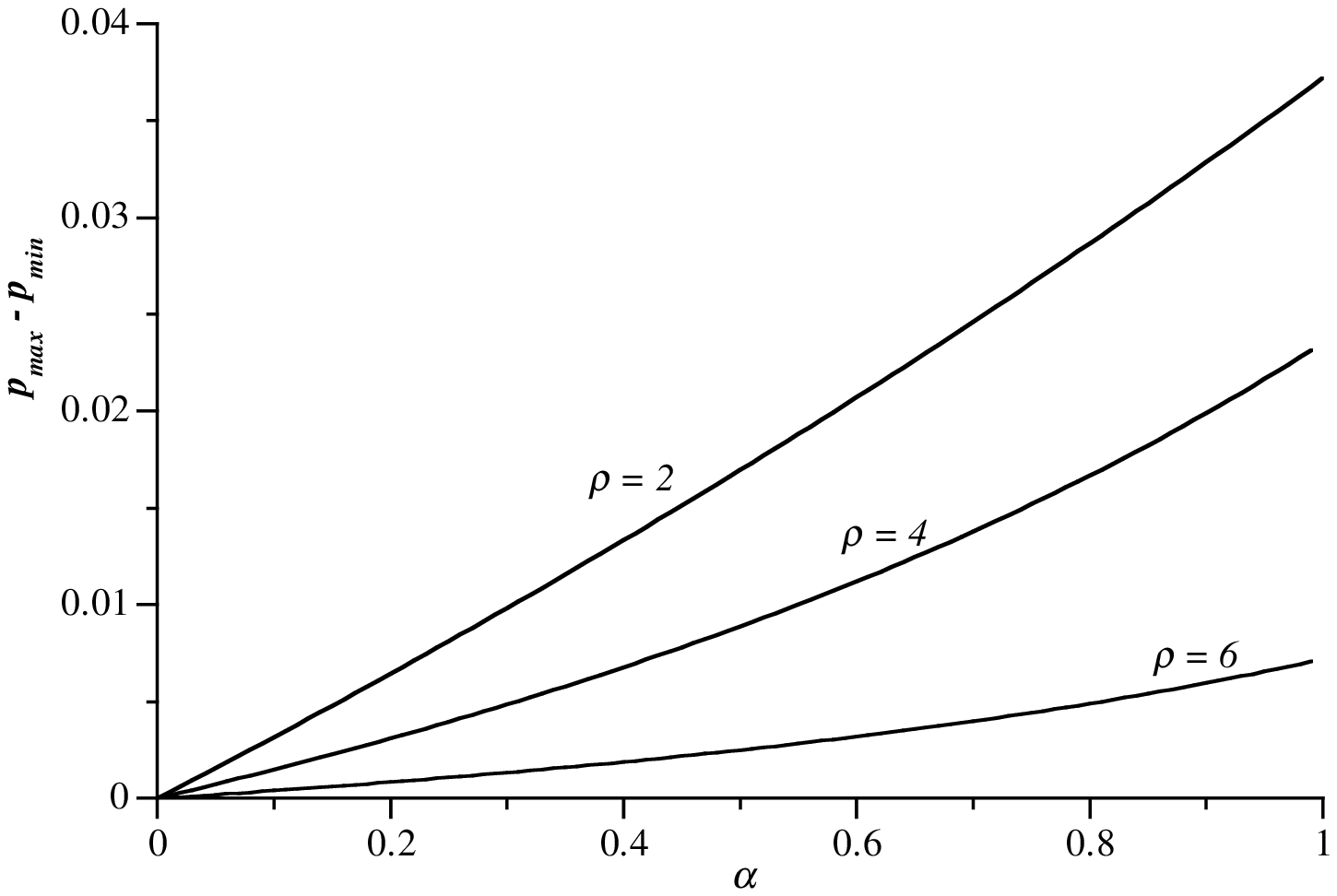,width=4.5in}}
\end{center}
\vspace*{13pt}
\fcaption{
Width of the plateaux $p_{max} - p_{min}$ ( (\ref{pmin}) - (\ref{pmax}) ) 
as function of $\a$ at different $\r = 2 , 4 , 6$.
}
\end{figure}
  
It is interesting to observe
that the plateaux disappear ($p_{max} \to p_{min}$) in 
the Kelly's limit ($\a \to 0$) and 
$p_{max}$ is the value for which 
the Kelly's model reaches its maximum when $l = \hat{l}$.
Then we notice from (\ref{lambdazero}) that for this value of $ p $
the maximum Lyapunov exponent rescaled with the function 
\begin{equation}
\eta(\a,p)=\ln(\a + 2 (1 - \a) p)
\,\, . \label{etaeqn}
\end{equation}
is the same for every $\a$.

Finally we notice that 
the Lyapunov exponent computed at the optimal value $l=l_{max}$, 
rescaled with (\ref{etaeqn}), is quantitatively independent of $\a$,
see Fig. 5.

\begin{figure}
\vspace*{13pt}
\begin{center}
\mbox{\psfig{file=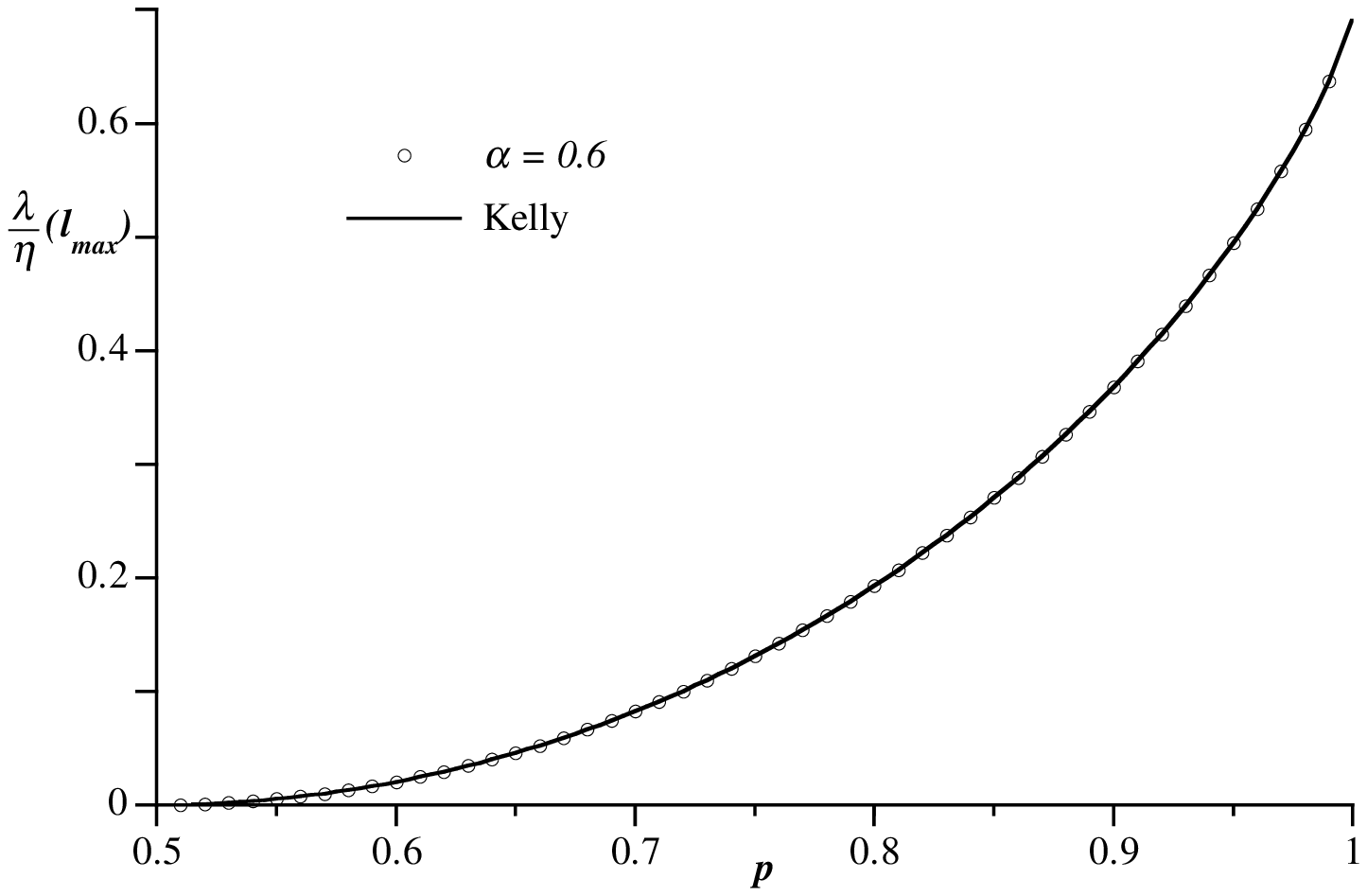,width=4.5in}}
\end{center}
\vspace*{13pt}
\fcaption{
Rescaled maximum Lyapunov exponent ${\l \over \eta}(l_{max})$ (full line)
as function of $p$  for $\a=0.6$, compared with the Kelly's case (circles).
}
\end{figure}

In addition the standard deviation $\Delta$ is proportional to 
the Lyapunov exponent when $l=l_{max}$.
This is an exact result, at any $\a$, when  $p = p_{max}$ and 
it is
qualitatively true for generic values of $p$, see Fig. 6.

\begin{figure}
\vspace*{13pt}
\begin{center}
\mbox{\psfig{file=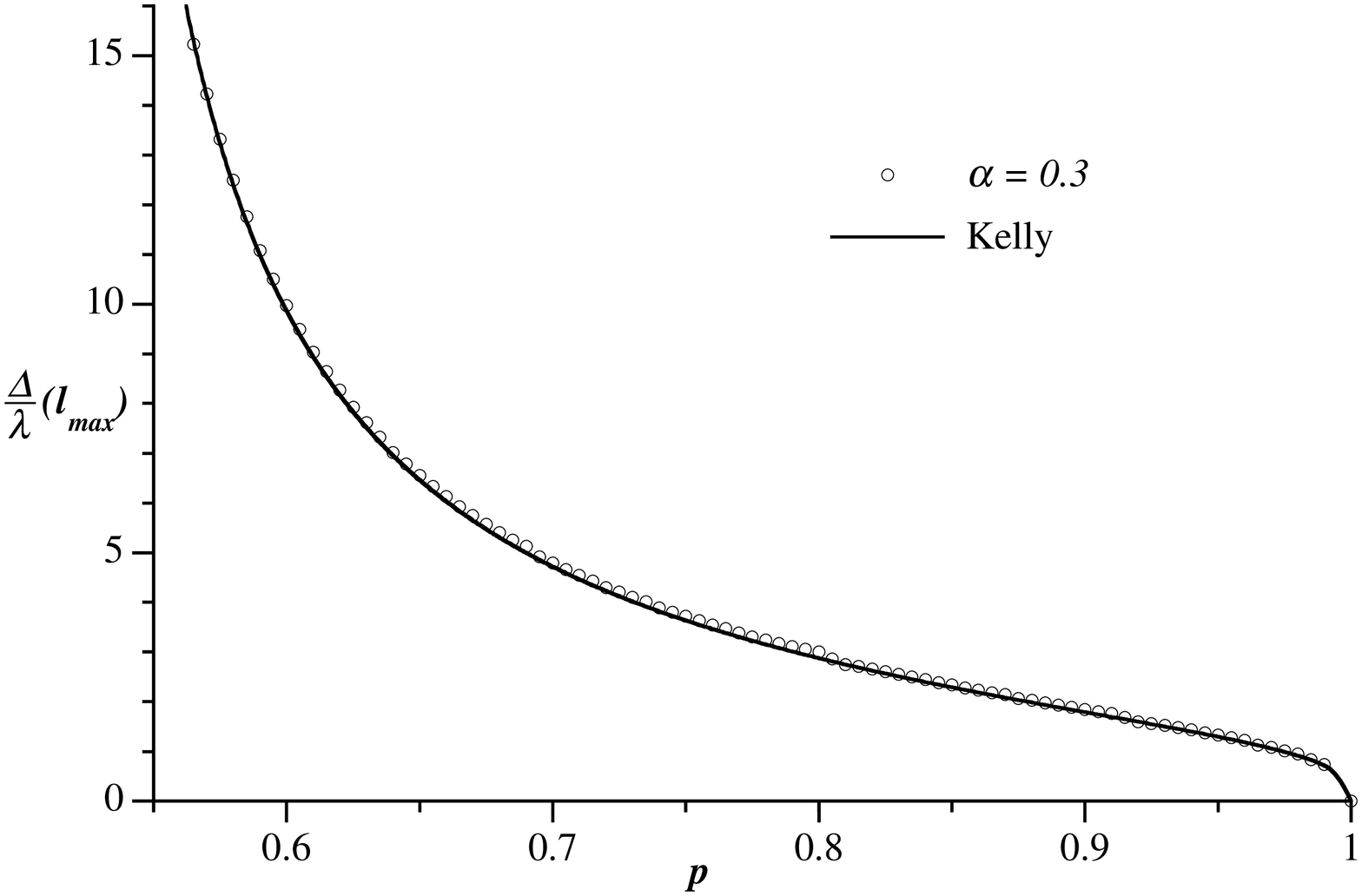,width=4.5in}}
\end{center}
\vspace*{13pt}
\fcaption{
Relative deviations ${ \Delta \over  \l } (l_{max})$
as function of $p$ for $\a=0.3$.
}
\end{figure}
   
Let us remind that a good parameter for quantifying the strength of the
fluctuation is the ratio $R = \Delta^2 / \l$. In fact,
in the log-normal distribution (\ref{lognorm}) both $\Delta^2$ and
$\l$ have the dimension of the inverse of time, and therefore
$\Delta^2 \over{\l}$ is time independent.
Since the $\Delta / \l$ is basically only function of $p$ and 
not of $\a$, one has

\begin{equation}
R = {\Delta^2 \over{\l}}  \simeq 
\left( {\Delta \over{\l} }\right)_{Kelly}^2 \l  \simeq
\left( {\Delta \over{\l} }\right)_{Kelly}^2 
{\eta \over{\eta_{Kelly}}} \l_{Kelly} =
{\eta \over{\eta_{Kelly}}} R_{Kelly} 
\le R_{Kelly}
\,\, . \label{fluctuations}
\end{equation}
i.e. a reduction of the relative fluctuations.

\section{Conclusions}
\noindent
In this paper we have considered a diversification of the 
portfolio between permanent goods and investments in a market.
The model is a non trivial modification of the Kelly's one where 
only a part of the capital is allowed to be invested on the 
market.
The investor keeps the remaining part as a security amount of money,
equal to a fraction $\a$ of the 
maximum capital owned in the past.
In this way the investor avoids the possibility of losing
almost all her capital due to a large fluctuation
as it can happen in the Kelly's case:
the parameter $\a$ can be thought as a measure of the
investor's prudence. 

The small fluctuations of the capital around the typical value
$W_0 \ e^{\l t}$ follow, at large $t$, a log-normal distribution
and therefore they are well described by 
the typical exponential growth (or Lyapunov exponent) $\l$ and 
the deviation $\Delta$ from this quantity.  
We give explicit analytical expressions 
for both these quantities. 
In particular we obtain a decreasing of the relative strength of the small
fluctuations (i.e. $\Delta^2/\l$).

An interesting feature of the model, from a mathematical point of view,
is that the Lyapunov exponent 
is a continuous but not differentiable function of the parameters.
This fact is particularly relevant when we look for 
the fraction $l$ 
of the allowed capital (i.e. the capital that can be 
invested in the market each time),
which maximizes $\l$.
We observe a devil's stairs like behaviour \cite{aubry} for $\r_{max}$
as a function of the probability $p$.
The existences of plateaux can be understood 
if one considers in more detail the $\l$ itself as a function of $l$ at fixed
$p$ and $\a$. The sizes of these plateaux can be computed,
as an example we derive the width of the largest ones. 

We have found that the Lyapunov exponent, when is rescaled by
a proper function of the parameters, and the ratio between
$\Delta$ and $\l$, show a similar behaviour of the Kelly's case:
the prudence's constraint we have introduced 
has basically the effect to rescale
the exponential growth and the relative strength of the small fluctuations
of the capital invested on the 
stock market according to equations (\ref{etaeqn}) and (\ref{fluctuations}).

The study of non commutative multiplicative models of the stock market has
the great advantage that they often can be analytically
treated. 
In generic cases one can use powerful systematic methods to obtain good
approximations \cite{cpv,serpal,pps}.
We belive they can be useful to understand problems
close to the reality, 
such as the ones where the investor does not decide only once 
the best strategy to follow, but can change at each step his mind
depending on the behaviour of the market.
It can be shown that a fixed strategy is not the best one 
when we introduce correlations between successive times 
or different {\it hedgings} for
prudent investors.

\nonumsection{Acknowledgment}
\noindent
We are very grateful to Erik Aurell 
for useful discussions.
We thank Yi-Cheng Zhang for a critical reading of the manuscript.
R.B. and M.S. acknowledge the Royal Institute of Technology of Stockholm
where part of this work has been developed
and in particular Duccio Fanelli for the warm hospitality.

\renewcommand{\theequation}{\Alph{appendixc}.\arabic{equation}}
\appendix

\noindent
In this appendix we derive the recursive formula (\ref{coeff})
for $C_n$. 
The random walk (\ref{modelbetween}) has a
positive drift ($p \ge {1\over2}$) and 
it ends as soon as the 
capital exceeds its initial value.
The total number of steps necessary for that is $N_n$,
where $n$ is the number of negative steps (i.e. defeats).
The stop is when $ n / N_n$ is smaller than
$1 / (\r +1)$ (with $\r \ge 1$),
so that 
$N_n = 1 + n + \left[ n \rho \right]$, 
where the square brackets mean integer part (see equation (\ref{timedur})).

The number of different ways $C_n$
can be computed noticing that 
the $n$ negative steps have to appear
before the $(N_{n-1})$-th step, in order to avoid a premature 
interruption of the process with only $n-1$ negative steps.
This yields to $\left(\matrix{N_{n-1} \cr n}\right)$
different combinations, but 
in this number are also included the cases 
of premature arrest with $r$ negative steps 
in the first $N_r$ time steps, with
$0 \le r \le n-2$.
Each of these cases leaves out an amount of
combinations equal to $\left(\matrix{N_{n-1}-N_r \cr n-r}\right) C_r$,
where the combinatorial factor comes from 
the different ways that the remaining $(n-r)$ negative steps
have to appear in the interval $[1+N_r , N_{n-1}]$.

It immediately follows the recursive formula (\ref{coeff})
\begin{eqnarray*}
C_n = & \left(\matrix{N_{n-1} \cr n}\right) - 
\sum_{r=0}^{n-2} \left(\matrix{N_{n-1}-N_r \cr n-r}\right) C_r =
\\ = & \left(\matrix{N_{n-1} -1 \cr n-1}\right) - 
\sum_{r=1}^{n-2} \left(\matrix{N_{n-1}-N_r \cr n-r}\right) C_r
\end{eqnarray*}
which holds for $n\ge 3$, while for $n=2$ one has
$$
C_2 = N_1 -1 
$$

The cases $n \le 1$ can be trivially derived:
$$
C_0 = C_1 = 1
$$

\nonumsection{References}

\end{document}